\documentclass{article}

\usepackage{cite, amsmath, amssymb}
\usepackage{url} 
\usepackage{graphicx}
\usepackage[affil-it]{authblk} 

\begin{document}
	
\title{Dynamical compensation in biological systems as a particular case of structural non-identifiability\footnote{An earlier version of this paper was submitted as correspondence to Molecular Systems Biology as a reply to \cite{karin} on December 9, 2016.}}
\author{\textbf{Alejandro F. Villaverde}\thanks{Email: \texttt{afvillaverde@iim.csic.es}} }
\author{\textbf{Julio R. Banga}}
\affil{Bioprocess Engineering Group, IIM-CSIC, Vigo, Spain}
\maketitle

\begin{abstract}
Dynamical compensation (DC) has been recently defined as the ability of a biological system to keep its output dynamics unchanged in the face of varying parameters. This concept is purported to describe a design principle that provides robustness to physiological circuits. Here we note the similitude between DC and Structural Identifiability (SI), and we argue that the former can be explained in terms of (lack of) the latter. We propose to exploit this fact by using currently existing tools for SI analysis to perform DC analysis. We demonstrate the feasibility of this approach with four physiological circuits, for which we confirm the correspondence between DC and lack of SI. We also warn that care should we taken when using an unidentifiable model to extract biological insight, since lack of SI can be the result of an inappropriate choice of model structure and therefore not necessarily a sign of biological robustness. 
\end{abstract}

\section{Dynamical compensation and structural identifiability}\label{sec:intro}

Some biological systems are capable of maintaining an approximatively constant output despite environmental fluctuations \cite{Barkal}. The ability to keep a constant steady state has been called exact adaptation. To account for the ability of preserving not only the steady state, but also the transient response, Karin et al. \cite{karin} have recently introduced the concept of dynamical compensation, which is defined as follows: 

``Consider a system with an input $u(t)$ and an output $y(t,s)$ such that $s > 0$ is a parameter of the system. The system is initially at steady state with $u(0) = 0$. Dynamical compensation (DC) with respect to $s$ is that for any input $u(t)$ and any (constant) $s$ the output of the system $y(t,s)$ does not depend on $s$. That is, for any $s_1$, $s_2$ and for any time-dependent input $u(t)$, $y(t,s_1) = y(t,s_2)$'' \cite{karin}.

The previous paragraph is similar to the classic definition of structural unidentifiability, which, using the same notation as in the DC definition, can be stated as follows (see e.g. \cite{ljung1994global,walter1997identification}): a parameter $s$ is structurally identifiable if it can be uniquely determined from the system output, that is, if for any $s_1$, $s_2$ it holds that $y(t,s_1) = y(t,s_2) \Leftrightarrow s_1 = s_2$. If this relationship does not hold for any $u(t)$, even in a small neighbourhood of $s$, the parameter is structurally unidentifiable. Thus, DC can be considered as a particular case of structural unidentifiability, with the additional requirement that the system is initially at steady state and with zero input. Due to the correspondence between the definitions of dynamical compensation and structural unidentifiability, we suggest that the latter concept can suffice to explain the phenomenon described by Karin and co-workers.

Structural identifiability (SI) is a well-established concept with a long history of applications in the biological sciences \cite{bellman1970structural}; indeed, it has historically been developed primarily by researchers working at the interface between biology and systems and control theory, long before the term ``systems biology'' became popular \cite{distefano2014dynamic}. The lack of structural identifiability is typical in overly complex models, i.e. those containing more parameters than can be supported by the evidence, even in the utopian case of perfect measurements. From its very origins, it was recognized that identifiability is concerned with the theoretical existence of unique solutions and therefore is, strictly speaking, a mathematical a priori problem \cite{jacquez1985numerical}. Furthermore, choosing a model structure is always based on a number of arbitrary decisions that can, nevertheless, have important consequences, so testing structural properties allows us to detect defects in the chosen model structure \cite{walter1997identification}. Therefore, a key question is whether we need to ensure structural identifiability in a model in order to extract insights about (and ultimately understand) the corresponding biological system. 

\section{DC and SI of four physiological circuits}

The four systems depicted in Figure 1 were used by Karin et al. \cite{karin} to illustrate the phenomenon of dynamical compensation. Their equations are provided in the original publication. Models of this type have been the subject of SI analyses in the past; for example, DiStefano \cite{distefano2014dynamic} reported identifiability issues in linear models of blood glucose control, such as the classic one by Bolie \cite{Bolie}, among others. From a broader perspective, the relevance of SI analysis in physiological modelling has recently been stressed e.g. by Janzen et al. \cite{janzen2016}. It is good news that much work has already been done in the field from the viewpoint of SI, and perhaps more important than the existence of publications analysing particular cases is the fact that there are general purpose methodologies available for performing that task. 

Some of the biological circuits studied by Karin et al. are non-linear (and, in the case of the $\beta$IG model of Fig. 1D, non-rational), which complicates their analyses. However, even in those cases there are a number of tools that can be readily applied. Using one such technique, the differential geometry approach implemented in the STRIKE-GOLDD toolbox \cite{villaverde2016}, we found that the parameters ($p, s$) of the two models exhibiting dynamical compensation--i.e. the hormone circuit of Fig. 1C and the $\beta$IG model of Fig. 1D--are structurally unidentifiable, while in the models that have exact adaptation but not dynamical compensation--i.e. the ones shown in Fig. 1A and 1B--those parameters are identifiable. Likewise, the parameters corresponding to endogenous glucose production ($u_0$) and insulin degradation ($\gamma$) in the $\beta$IG model are classified as structurally identifiable by STRIKE-GOLDD, as might be expected given that the model does not have DC to them. 

All these results are consistent with our claim that there is a correspondence between a model having DC with respect to a parameter and that parameter being structurally unidentifiable. We remark that STRIKE-GOLDD performs a symbolic analysis, which does not require knowledge of the numeric values of the parameters. Thus, unlike in \cite{karin}, these results were obtained without simulating the models. 
The code and instructions for running these analyses can be found in \url{https://sites.google.com/site/strikegolddtoolbox/dc}.

\begin{figure*}[t]
	\begin{center}
		\includegraphics[width=0.9\linewidth]{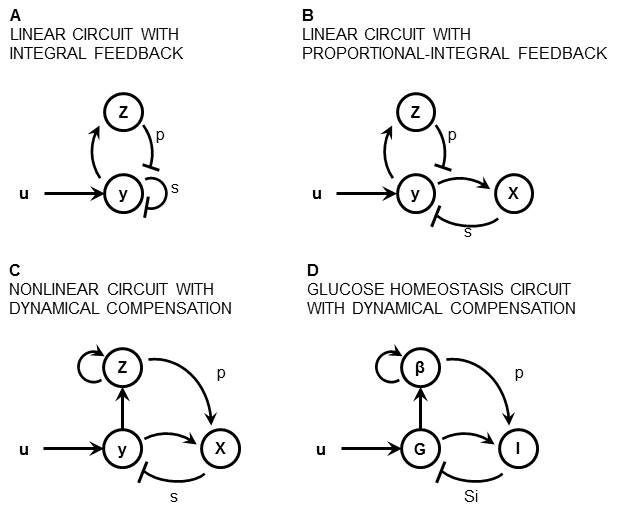}
		\caption{Physiological systems used as case studies. The four circuits shown here are deterministic dynamic models described by ordinary differential equations (ODEs), which can be found in \cite{karin}.}  
	\end{center}
	\label{fig:4circuits}
\end{figure*} 

\section{Discussion}

The fact that a model is unidentifiable is important because, after five decades of research, it is now well understood that lack of structural identifiability can be the result of choosing an inappropriate model structure and therefore not necessarily a sign of biological robustness. When understood as a modelling artefact, it can be avoided by e.g. (i) reformulating the model, or by (ii) increasing the number and type of measurements. Strategy (i) has actually been noted by Karin and co-workers, who showed (see equations 4-6 of the original paper) that it is possible to remove the unidentifiable parameters from the model by transforming the state variables. 

Some considerations about sensitivity, identifiability, and robustness will help clarify this point. If the state of a system changes very little or not at all when a parameter varies, the system is said to be robust or insensitive to variations in that parameter \cite{distefano2014dynamic}. Biological robustness is revealed by low sensitivity of a model output to a particular parameter. Speaking in terms of identifiability, this situation corresponds to poor \textit{practical} (or \textit{numerical}) \textit{identifiability}: although the value of the parameter has some influence on the model output, its effect is too small to allow for its precise determination due to limitations in data quantity and/or quality \cite{jacquez1985numerical,walter1997identification}. In contrast, a sensitivity of the model output to a parameter that is exactly zero—as implied by DC—corresponds to lack of \textit{structural identifiability}. In this case, the value of the parameter has no influence at all on the model output: it can be changed e.g. a thousand-fold without producing any effect. This ``unreasonable elasticity'' indicates that the parameter is not meaningful; ideally, it should be removed and the model should be modified.

Interestingly, structural identifiability has close links with other systemic properties, namely observability and controllability, which provide important information about a dynamic model. Observability determines whether it is possible to reconstruct the internal state of a model by observing its output, while controllability refers to the possibility of controlling its state by manipulating its input. Identifiability and observability are tightly related; in fact, SI analysis can be recast as observability analysis by considering the model parameters as constant state variables. In turn, observability and controllability are usually considered as dual concepts. These three distinct but related properties, which characterize important aspects of biological systems, can be analysed for general nonlinear ODE models using differential geometric approaches, which are described e.g. in \cite{Vidyasagar1993}.

The introduction of the concept of dynamical compensation has been motivated by the study of adaptation in physiological circuits. It has thus been developed in a different context than structural identifiability, and it may provide a fresh perspective to an old problem, complementing existing approaches. We believe that it is beneficial to draw a connection between the newly coined concept of dynamical compensation and the already existing theory on identifiability. We see this case as another example of the gains that can be obtained by exchanging more notes among the different communities working in systems biology, which we have advocated elsewhere \cite{villaverde2014reverse}.

\end{document}